
\input phyzzx
\hfuzz 20pt
\font\mybb=msbm10 at 10pt

\def\Bbb#1{\hbox{\mybb#1}}

\def\bR{\Bbb {R}}
\def\bE{\Bbb {E}}

\def\bfomega{\omega\kern-7.0pt \omega}

\magnification=900



\def\C{\mkern1mu\raise2.2pt
\hbox{$\scriptscriptstyle|$}\mkern-7mu{\rm C}}

\def\pd{\partial_}
\def\pu{\partial^}

\def\m{\mu}
\def\n{\nu}
\def\l{\lambda}

\def\p{\rho}
\def\s{\sigma}
\def\t{\tau}

\def\f{\phi}
\def\d{\delta}
\def\e{\epsilon}

\def\cN {{\cal {N}}}


\font\mybb=msbm10 at 12pt

\def\Bbb#1{\hbox{\mybb#1}}

\def\bR{\Bbb {R}}
\def\bE{\Bbb {E}}

\def\bfomega{\omega\kern-7.0pt \omega}


\REF\gibr {G.W.Gibbons \& P.J.Ruback,{\sl
The motion of extreme Reissner-Nordstr\"om black
holes in the low velocity limit}, Phys. Rev.
Lett. {\bf 57} (1986) 1492.}
\REF\fere {R.C.Ferrell \& D.M.Eardley,{\sl
Slow motion scattering and coalescence of
maximally charged black holes},Phys. Rev.
Lett. {\bf 59} (1987) 1617.}
\REF\shiraishi{K. Shiraishi, {\sl Moduli 
Space Metric for Maximally-Charged
Dilaton Black Holes}, Nucl. Phys. {\bf B402} (1993) 399.}
\REF\coles { R.A.Coles and G.Papadopoulos,{\sl The
Geometry of one-dimensional supersymmetric
nonlinear sigma models}, Class. Quantum Grav.
{\bf 7} (1990) 427.}
\REF\hull{C.M. Hull, {\sl The geometry of 
supersymmetric quantum mechanics}, hep-th/9910028.} 
\REF\gary{G.W. Gibbons, G. Papadopoulos 
and K. Stelle, {\sl HKT and OKT
Geometries on Soliton Black Hole 
Moduli Spaces}, Nucl. Phys. {B508} (1997) 623,
hep-th/9706207.}
\REF\micha{J. Michelson and 
A. Strominger, {\sl Superconformal 
Multi-Black Hole
Quantum Mechanics}, HUTP-99/A047, 
hep-th/9908044.\break
R. Britto-Pacumio, J. Michelson, 
A. Strominger and A. Volovich,
{\sl Lectures on superconformal 
quantum mechanics and multi-black hole
moduli spaces}, hep-th/9911066.}
\REF\twist{P.S. Howe and 
G. Papadopoulos, {\sl Twistor Spaces
for HKT Manifolds}, Phys. 
Lett. {\bf B379} (1996)80, hep-th/9602108.}
\REF\gutpap{J. Gutowski and 
G. Papadopoulos, {\sl The dynamics 
of very special black holes}, 
hep-th/9910022 .}
\REF\mss{A. Maloney, M. Spradlin 
and A. Strominger, {\sl Superconformal
Multi-Black Hole Moduli Spaces 
in Four Dimensions}, hep-th/9911001.}
\REF\sabralust{K. Behrndt, D. Lust, W. A. Sabra, 
{\sl  Stationary solutions of N=2 supergravity}, 
 Nucl.Phys. {\bf B510} (1998) 264; hep-th/9705169.} 
\REF\gppkt{G. Papadopoulos and P.K. Townsend, 
{\sl Intersecting M-Branes}, 
Phys.\ Lett.\ {\bf B380} (1996) 273; hep-th/9603087.}
\REF\arkkleb{I.R. Klebanov and 
A.A. Tseytlin, {\sl Intersecting M-branes
as four-Dimensional Black Holes}, hep-th/964166.}
\REF\arkady{A. Tseytlin, {\sl Composite Black 
Holes in String Theory}, hep-th/9608044.}
\REF\fin{V. Balasubramanian, F. Larsen and R. G. Leigh,
{\sl Branes at angles and black holes}, 
Phys. Rev. {\bf D57} (1998) 3509; hep-th/9704143.} 
\REF\strom{A. Strominger 
and C. Vafa, {\sl Macroscopic Origin of the
Bekenstein-Hawking Entropy}, Phys. Lett. {\bf B379} 
(1996) 99, hep-th/9601029.}
\REF\callan{ C. G. Callan, Jr., J. M. Maldacena, 
{\sl D-brane approach to black hole quantum mechanics}, 
 Nucl.Phys.{\bf B472} (1996) 591; hep-th/9602043.}
 \REF\revi{ P.K. Townsend, {\sl The 
 eleven-dimensional supermembrane revisited}, 
Phys.Lett. {\bf B350} (1995) 184; hep-th/9501068.} 
\REF\stelle{
M.J. Duff, K.S. Stelle, {\sl Multi-membrane 
solutions of d = 11 supergravity},
Phys.Lett. {\bf B253} (1991) 113.}
\REF\guven{R. Guven, {\sl Black p-brane 
solutions of D=11 supergravity theory},
Phys. Lett. {\bf 276B} (1992) 49.}
\REF\hs{G.T. Horowitz and A. Strominger, {\sl Black 
strings and p-branes},
Nucl. Phys. {\bf B360} (1991) 197.}
\REF\dl{M.J. Duff and J.X. Lu, {\sl The selfdual 
type IIB superthreebrane },
 Phys. Lett. {\bf B273} (1991) 409.}
\REF\dgh{A. Dabholkar, G.W. Gibbons, 
J.A. Harvey and F. Ruiz-Ruiz, 
{\sl Superstrings and solitons}, Nucl. 
Phys. {\bf B340} (1990) 33.}
\REF\chs{
C. G. Callan, Jr., J. A. Harvey,
 A. Strominger, {\sl Worldbrane actions 
for string solitons},
 Nucl.Phys. {\bf B367} (1991)60.}
 \REF\arkadyb{A.A.~Tseytlin,
{\sl Harmonic superpositions of M-branes},
Nucl.\ Phys.\ {\bf B475} (1996) 149;
hep-th/9604035.}
\REF\john{
C. V. Johnson, A. W. Peet and 
J. Polchinski, {\sl Gauge Theory and 
the Excision of Repulson 
Singularities};  hep-th/9911161. } 
\REF\town{M. G\"unaydin, G. Sierra and 
P.K. Townsend,  {\sl Exceptional
Supergravity Theories and the 
Magic Square}, Phys. Lett. {\bf B133}
(1983) 72; {\sl The Geometry of 
N=2 Maxwell-Einstein
Supergravity and Jordan Algebras}, 
Nucl. Phys. {\bf B242} (1984) 244.}
\REF\wit{B. de Wit and A. van Proeyen, {\sl Broken 
Sigma Model Isometries in
Very Special Geometry}, Phys. Lett. 
{\bf 293B} (1992) 95, hep-th/9207091.}
\REF\sabra{W.A. Sabra, {\sl General BPS Black Holes
in Five Dimensions}, Mod. Phys. Lett. {\bf A 13} 
(1998) 239, hep-th/9708103; 
A. Chamseddine and W.A. Sabra, 
{\sl Metrics admitting Killing 
spinors in five dimensions}, 
Phys. Lett. {\bf B426} (1998) 36.}
\REF\sabratwo{ A. Chamseddine and 
W.A. Sabra, {\sl Calabi-Yau black holes
and enhancement of supersymmetry 
in five dimensions},  Phys. Lett. 
{\bf B460} (1999) 63; hep-th/9903046.}
\REF\mirone{M. Cvetic and  D. Youm,
{\sl Dyonic BPS Saturated Black Holes 
of Heterotic String on a Six Torus},
 Phys.Rev. {\bf D53} (1996) 584; hep-th/9507090.} 
 \REF\mirtwo{M. Cvetic and A. Tseytlin,  
{\sl General Class of BPS Saturated 
Dyonic Black Holes as Exact Superstring
Solutions},
 Phys.Lett. {\bf B366} (1996) 95, hep-th/9510097;
{\sl Solitonic Strings and BPS Saturated Dyonic Black Holes}
 Phys.Rev. {\bf D53} (1996) 5619, 
 Erratum-ibid. {\bf D55} (1997) 3907,
 hep-th/9512031.} 
\REF\kalloshc{A. Chou, R. Kallosh, 
J. Rahmfeld, S-J. Ray, M. Shmakova
and W. K. Wong, {\sl Critical 
Points and Phase transitions
in 5-D compactifications of M-theory}, Nucl. Phys.
 {\bf B508} (1997) 147; hep-th/9704142.}
\REF\cfgk{A.H. Chamseddine, S. Ferrara, 
G.W. Gibbons and R. Kallosh,
{\sl Enhancement of supersymmetry 
near black hole horizons}, Phys. Rev.
{\bf D55} (1997) 3647; hep-th/9610155.}
\REF\gwgaa{ G.W. Gibbons, {\sl Antigravitating 
black hole 
solitons with scalar hair in N = 4 supergravity}, 
 Nucl.Phys. {\bf B207} (1982) 337.} 
\REF\renatapeet{.
 R. Kallosh and A. Peet, {\sl Dilaton 
 black holes near the horizon},
  Phys.Rev. {\bf D46} (1992)5223; hep-th/9209116} 
\REF\gwgpktb{G.W. Gibbons and P.K. 
Townsend, {\sl Vacuum interpolation in 
supergravity via super p-branes},  
Phys.Rev.Lett. {\bf 71} 
(1993)3754; hep-th/9307049.}
\REF\stu{I. Antoniadis, S. Ferrara 
and T.R. Taylor, {\sl N=2
Heterotic Superstrings and its 
Dual Theory in Five Dimensions},
 Nucl. Phys. {\bf B460} (1996)
489; hep-th/9511108.}
\REF\sabrac{I. Gauda, S. Mahapatra, 
T. Mohaupt and W. A. Sabra, {\sl Black Holes and
Flop Transitions in M-theory on 
Calabi-Yau Threefolds}, Class. Quantum
Grav. {\bf 16} (1999) 419; hep-th/9807014.}
\REF\kalloshb{R. Kallosh, A. Linde 
and M. Shmakova, {\sl Supersymmetric
Multiple Basin Attractors}, hep-th/9910021.}
\REF\poon{G. Grantcharov and Y.S. Poon, 
{\sl Geometry of hyper-K\"ahler
Connections with Torsion}, math.dg/9908015.}
\REF\michb{J. Michelson and A. Strominger, {\sl The 
geometry of (Super)conformal
Quantum Mechanics}, HUTP-99/A045, hep-th/9907191.}
\REF\george{G. Papadopoulos, {\sl Conformal 
and Superconformal Mechanics};
hep-th/0002007.}
\REF\spgeo{B. de Wit and A. van Proeyen, 
{\sl Potential and Symmetries
of General Gauged N=2 Supergravity-Yang-Mills Models}, 
Nucl. Phys. {\bf B245}
(1984) 89;\break
E. Cremmer, C. Kounnas, A. van Proeyen, 
J.P. Derendinger,
 S. Ferrara,
B. de Wit and L. Girardello, {\sl Vector Multiplets 
Coupled to N=2 Supergravity: Superhiggs
 Effect, Flat
Potentials and Geometric Structure}
 Nucl.Phys.{\bf B250} (1985)385.}  
\REF\tesc{G. Papadopoulos and 
A. Teschendorff, {\sl Multi-angle
Five-Brane Intersections}, Phys. Lett. {\bf B443} (1998) 
159, hep-th/9806191;
{\sl Grassmannians, Calibrations and
 Five-Brane Intersections}, hep-th/9811034.}
 \REF\mirjam{
 M. Cvetic and
C. M. Hull, {\sl Black Holes and U Duality},
 Nucl.Phys. {\bf B480} (1996)296; hep-th/9606193.} 
 \REF\fre{M. Bertolini, P. Fre and 
 M. Trigiante, {\sl The generating 
solution of regular N=8 BPS black holes}, 
Class.Quant.Grav. {\bf 16}
(1999)2987, hep-th/9905143.\break
M. Bertolini and M. Trigiante, {\sl
Regular R-R and NS-NS BPS Black Holes}, hep-th/9910237;
{\sl Regular BPS Black Holes: Macroscopic 
and Microscopic Description of
the Generating solution}, hep-th/0002191.} 
\REF\pol{J. Polchinski, {\sl Tasi 
lectures on D-branes}, hep-th/9611050.}
\REF\costa{M.S. Costa and G. Papadopoulos, 
{\sl Superstring dualities and
 p-brane bound states}, Nucl.Phys. 
 {\bf B510} (1998) 217; hep-th/9612204.}
\REF\fubini{V. de Alfaro, 
S. Fubini and G. Furlan, {\sl Conformal
Invariance in Quantum Mechanics}, 
Nuovo. Cim. {\bf 34A} (1976) 569.}
\REF\peet{P. Claus, M. Derix, 
R. Kallosh, J. Kumar, P.K. Townsend,
A. van Proeyen, {\sl Black Holes and 
Superconformal Mechanics}, Phys. Rev.
Lett. {\bf 81} (1998) 4553; hep-th/9804177.}
\REF\ggpkt{G.W. Gibbons and P.K. Townsend, 
{\sl  Black Holes and Calogero Models},
 Phys.Lett. {\bf B454} (1999) 187; hep-th/9812034.}


\date{February 2000}
\titlepage
\title{Moduli Spaces for Four- and Five- Dimensional  Black Holes}
\author{J. Gutowski}
\address{Center for Mathematical Sciences, 
University of Cambridge,\break
Wilberforce Road, Cambridge CB3 0WA.}
\andauthor{G. Papadopoulos}
\address{Department of Mathematics, 
King's College London,\break
Strand, London WC2R 2LS.}
 
 \abstract{We propose a universal 
 expression for the moduli metric
of a class of four- and five-dimensional 
black holes which preserve at least
four  supersymmetries. These include 
the black holes that are
associated with various intersecting 
branes in ten and eleven dimensions,
the electrically charged black holes of
 N=2 D=5 and N=2 D=4 supergravities with
any number of vector multiplets,  and  
dyonic black holes of N=2 D=4 supergravity.
The moduli 
 metric of electrically charged 
 N=2 D=4 black holes coupled to any
 number of vector multiplets is 
 explicitly computed.   We also
investigate the superconformal 
symmetries of the black hole moduli spaces
for small black hole separations.}

\endpage

\pagenumber=1




\chapter{Introduction}

In the past few years there has been much 
interest in investigating the geometry
of the moduli spaces of various 
supersymmetric black hole solutions
of supergravity theories following some earlier work in
 [\gibr, \fere, \shiraishi]. Supersymmetric
 black hole solutions are thought as
the solitons of supergravity and 
so provide a macroscopic description 
of the solitons of strings and M-theory. As such
one can investigate their moduli 
spaces in analogy with  similar investigations
of the moduli spaces of BPS
 monopoles in the context of Yang-Mills theory.
However unlike the case of BPS 
monopoles, the geometry of the moduli
space of various black holes 
is related to that of the target space
of supersymmetric sigma models in one-dimension 
with {\sl non-vanishing torsion} 
[\coles]; for more recent work
on the geometry of one-dimensional 
sigma models see [\hull].
This has been first established in 
[\gary] for a class of D=5 black holes
that preserve 1/4 of the maximal 
supersymmetry and later extended  for the
electric black holes of D=5 N=2 
supergravity coupled  to the graviphoton [\micha].
 In the former case, the 
 geometry of black hole moduli space is
 strong HKT while in the latter 
 is weak HKT  [\twist].
 Later it was found that the
  moduli space of electrically
   charged black holes
 of N=2 D=5 supergravity with 
 any number of vector multiplets
 is again weak HKT [\gutpap]. 
More recently,  the  moduli space of 
(four-dimensional) Reissner-Nordstr\"om
 black holes 
was investigated [\mss] and again 
it was confirmed that its geometry
is related to that  of a class 
of one-dimensional
sigma models which in addition to 
some bosonic multiplets also contain
 fermionic ones [\coles]. A common 
 characteristic of all the above cases is
that the geometry on the moduli space 
of these black holes is determined
by a scalar function, a \lq moduli potential'. 
The authors of [\micha, \mss] also
investigated the symmetries of the
 moduli spaces of the N=2 D=5 graviphoton 
and the Reissner-Nordstr\"om black 
holes for small black hole separations
and they found that they exhibit a $D(2,1;0)$
 superconformal symmetry.

In this paper, we propose a moduli 
metric for a class of black holes
in four and five dimensions that 
preserve at least four supersymmetry charges
of the underlying supergravity 
theory. Typically, we consider black holes
of  maximal supergravities  or black holes
of N=2 supergravities in four and five dimensions.
To be specific, if the metric of supersymmetric
black holes is given by
$$
ds^2=-A^2 dt^2+B^2 d{\bf x} d{\bf x}
\eqn\bhmetric
$$
then the moduli potential $\mu$ is
$$
\mu=\int d^{D-1}x\, A^{-2} B^2\ , 
\eqn\unipot
$$
where $D=5$ or $D=4$.
The integration is over the 
spatial directions of the black hole
with respect to the Euclidean metric.
It is assumed that the solution
is perturbed within the appropriate 
supergravity theory and the only
moduli parameters are the positions of the black holes. 
The above choice of the moduli potential 
is independent from the choice 
of frame of the associated
supergravity theory. One novel 
property of our expression for the
moduli potential is that it includes 
all the examples of known black hole moduli
potentials that have been 
computed so far. We also explicitly
compute  the moduli metric
of the electrically charged 
black hole solutions of N=2 D=4 
supergravity [\sabralust] in section eight and we find that 
it is again given by \unipot. 
In addition, we verify that the 
effective action associated with
the moduli potential \unipot\ of a certain
class of black holes  exhibits
 the
expected superconformal 
symmetries for small black hole separations.
The metric and torsion on the 
black hole moduli space associated
with the moduli potential \unipot\ 
are given in sections six and seven.

We first apply our formula to four- 
and five-dimensional 
black holes that can be constructed by
reducing intersecting brane 
configurations from ten and eleven dimensions.
We show that the moduli potential 
is invariant under the T-and S-dualities
of type II strings and the choice 
of frame of the  supergravity
theories. Moreover, \unipot\ can be 
partially motivated by the universality of the
ratio of the conformal factor of 
the transverse directions of the branes
modulo that of the worldvolume directions.
Since the moduli space of a class 
of electrically charged black holes
has been explicitly computed, if one 
assumes that the moduli metric is
S- and T-duality invariant, then 
one can establish \unipot.
Another application of the moduli 
potential \unipot\ is in establishing
the moduli metric of intersecting 
brane configurations as  will be explained
in section three.

We next apply our formula \unipot\ to a 
class of four- and five-dimensional
black holes associated with N=2 D=5 
supergravity with any number of vector
multiplets. The moduli potential 
of the electrically charged black holes
of this theory has been explicitly
 computed and agrees with \unipot. This
includes the case of five-dimensional
 black holes coupled to the
graviphoton and investigated in [\micha]. 
In four dimensions our formula
agrees with the moduli potential 
computed from the moduli metric
of (four-dimensional) Reissner-Nordstr\"om 
black holes. Next we  apply \unipot\
to give the moduli potential of the dyonic 
four-dimensional black holes that arise
from the reduction of the string 
solutions of N=2 D=5 supergravity
superposed with a pp-wave and the 
electrically charged N=2 D=5 black holes
in the background of a KK-monopole. 
We also verify with an explicit computation
that the moduli potential of the 
four-dimensional electrically
charged black holes of N=2 D=4 
supergravity is again given by \unipot.
We remark that the above mentioned dyonic black holes
are dual to these electric ones. 
Finally, we investigate the
 supersymmetric and superconformal properties
 of the effective
actions of all the above black holes. 
We find that for the class of such
black holes which have regular 
horizons the effective action 
exhibits $D(2,1;0)$ superconformal 
symmetry at small black hole
separations.

This paper has been organized as 
follows: In section two, we
describe the moduli potential of 
four- and five-dimensional black holes
that arise from brane intersections
 in ten and eleven dimensions.
In section three, we provide 
 evidence in support of
\unipot\ using duality. In section four, 
we give the four-dimensional dyonic
black hole solutions which are reductions 
of the string solutions superposed
with a pp-wave and the electrically 
charged solutions superposed with a KK-monopole
 of  N=2 D=5
supergravity. In section five, we 
apply our formula to give the moduli
potential of all the above  black 
holes and as an example we present the
moduli potential of black holes 
associated with 
the STU model. In section six, 
we construct the effective theory
of five-dimensional black holes 
and examine its superconformal properties.
In section seven, we construct 
the effective theory of four-dimensional
black holes and examine its
 superconformal properties. In section eight,
we compute the moduli
 metric of the electrically
charged four-dimensional 
black holes of N=2 supergravity
coupled to any number of
 vector multiplets and in section 
nine we give our conclusions.


\chapter{Black Holes in four and five dimensions }

A large class of four- and five-dimensional 
black holes\foot{We use the term
black holes to characterize all static solutions 
of supergravity
which are asymptotically flat. In particular, we do not
 require for the solutions
to have an  event horizon.}
can be constructed by appropriately reducing 
intersecting brane configurations
of strings and M-theory [\gppkt]. 
This has been widely 
explored in the literature 
[\gppkt, \arkkleb, \arkady, \fin]. These in particular
include the black holes that 
have been used to perform a microscopic
computation of the 
Bekenstein-Hawking entropy in [\strom, \callan].
It has been observed in [\arkady] 
that the metric of such black 
hole solutions can be expressed
 as
$$
ds^2=-\lambda^{D-3} dt^2+\lambda^{-1} ds^2(\bR^{D-1})
\eqn\bbh
$$
where
$$
\lambda=(\Pi^n_{I=1}H_I )^{-{1\over D-2}}\ ,
\eqn\ina
$$
and
$$
\Pi^n_{I=1}H_I=H_1\dots H_n\ .
\eqn\inb
$$
The functions $H_I$ are harmonic  in $\bR^{D-1}$, i.e.
$$
H_I=h_I+\sum_A {\lambda_{IA}\over |x-y_{IA}|^{D-3}}\ .
\eqn\inc
$$
The constants $\{h_I; I=1,\dots, n\}$ 
can be related to the
asymptotic values of the various 
scalars of the supergravity theory, 
$\{ \lambda_{IA}; A=1,\dots, N_I; I=1,\dots,n\}$ 
are the black hole charges and 
$\{ y_{IA}; A=1,\dots, N_I; I=1,\dots,n\}$ 
are the black hole positions.
For five-dimensional black holes 
(D=5) $n\leq 3$ and for four-dimensional
black holes (D=4) $n\leq 4$.
A subclass of such black holes are those for which
the positions of the harmonic functions coincide, i.e.
$$
y_{IA}=y_{JA}
\eqn\ind
$$
for $I\not=J$.
If in addition $n=3$ for $D=5$
 or $n=4$ for $D=4$, then these
black holes have regular horizons.

The black holes that are described
 by the metric \bbh\ carry electric
or magnetic or both electric and 
magnetic charges. The origin
of these charges can be traced in 
their interpretation as intersecting
branes in ten or eleven dimensions.  
In broad terms if the black hole
is associated with M-2-branes and 
pp-waves, then it is electrically charged but if
it is associated with M-5-branes 
and KK-monopoles, then it is magnetically charged.
There are also dyonic black holes 
that are associated with both electric and magnetic
branes. The Maxwell fields that 
the above black holes couple to are either
KK-vectors due to the reduction 
or they are associated to the various
brane field strengths in ten 
and eleven dimensions. In some cases,
it is possible to take linear
 combinations of the Maxwell fields
such that it can appear that a 
black hole couples to fewer Maxwell
fields than it may be expected.  
This mostly arises when we set the
various harmonic functions that
 the black holes depend on to be equal.
Since, we shall not use the 
explicit expression of the Maxwell fields
of the solutions in what follows we shall neglect them.

Using the expression for the moduli  
potential proposed in the
introduction, we find that
$$
\mu=\int d^{D-1}x\, \lambda^{2-D}\, 
\eqn\inaa
$$
or equivalently
$$
\mu=\int d^{D-1}x\, \Pi^n_{I=1}H_I\ .
\eqn\inab
$$

In the five-dimensional case (D=5),
 for $n=1$ the moduli space is flat, for
$n=2$ the moduli space is strong HKT 
and for $n=3$ the moduli space is weak HKT.
Such solutions preserve $1/2$, $1/4$ 
and $1/8$ of supersymmetry, respectively.
Moreover they are associated with 
configurations in strings and M-theory
that involve one brane, two branes 
and three branes, respectively.
The general case is that with three 
harmonic functions since all the
others can be derived from it by 
setting one or more of the harmonic functions
to be constant. Another 
simplification of the $n=3$ case is to set
all the harmonic functions to be equal, i.e.
$$
H=H_1=H_2=H_3\ .
\eqn\inbb
$$
In that case, the moduli potential becomes
$$
\mu=\int\, d^4x\, H^3\ .
\eqn\threesame
$$
This moduli potential is the 
same as that of the black holes 
of N=2 supergravity coupled to 
the graviphoton and derived in [\micha].
In fact, the graviphoton black 
hole can be constructed by reducing the
M-brane configuration of three 
intersecting M-2-branes on a 0-brane
with all three harmonic functions associated with each 
M-2-brane set to be equal [\gppkt].
In this case, the moduli 
potential \threesame\ has been verified 
by an explicit calculation. 
For the rest of the cases, we shall provide
an argument in the next section.

The superconformal properties of 
the moduli space \threesame\ for
small black hole separation  are
 the same as those  of graviphoton black holes
investigated in [\micha].  A more 
general case arises whenever we choose the positions
of the harmonic functions to be 
the same but the asymptotic values of the 
scalars and the charges to be different, i.e.
$$
H_I=h_I+\sum_A {\lambda_{IA}\over |x-y_{A}|^{D-3}}\ .
\eqn\harmeq
$$
The superconformal properties
of these black holes will be 
investigated in section six.

In the four-dimensional case (D=4), 
for $n=1$ the moduli 
space is flat and for $n=4$
the geometry on the moduli space 
generalizes that on the moduli space 
of the Reissner-Nordstr\"om
black holes. The rest of the 
cases are new. For $n=1$ the solutions preserve 
$1/2$, for $n=2$ the
solutions preserve $1/4$, for $n=3$ and $n=4$ the 
solutions preserve $1/8$ of the maximal
supersymmetry,
respectively. As in the five-dimensional
case, the most general case 
arises whenever $n=4$ since all the other
cases can be derived by setting 
one or more harmonic functions to be constant.
Another simplification of the $n=4$ case is to set
all the harmonic functions to be equal, i.e.
$$
H=H_1=H_2=H_3=H_4\ .
\eqn\indd
$$
In that case, the moduli potential becomes
$$
\mu=\int\, d^3x\, H^4\ .
\eqn\foursame
$$

The geometry on the moduli space then is 
that of the Reissner-Nordstr\"om
black holes [\mss]. So the superconformal 
properties of the effective theory
for small black hole separation are the 
same as those of the Reissner-Nordstr\"om
black holes.
A more general case arises whenever 
we choose the positions
of the harmonic functions to be the 
same but the asymptotic values of the 
scalars and the charges to be 
different as in \harmeq. The superconformal properties
of these black holes will be 
investigated in  section seven.


\chapter{Moduli Spaces from Intersecting Branes}

It is well known that all of the
 supersymmetric black holes
in four- and five-dimensions can be
 constructed by reducing
intersecting M-brane configurations  
possibly superposed
with a pp-wave or a KK-monopole. The 
two latter configurations
reduce to a D-0-brane and a D-6-brane
 upon compactification to
ten dimensions on $S^1$, respectively [\revi].
So one may expect to understand the 
expression of the moduli
potential by investigating the brane 
solutions in eleven and ten dimensions.
We shall argue that indeed some of
 the features of the
proposed moduli potential of the
 four- and five-dimensional black holes
have their origin in the form of the
 brane solutions in ten and eleven
dimensions. But there are also some puzzles.

The first observation toward this 
concerns the ratio of the
components of the metric of all brane
 solutions in ten and eleven dimensions.
To be specific recall that the  
spacetime metrics 
of the various branes are as follows:
The metric of the M-2-brane 
[\stelle] and the M-5-brane [\guven]
are
$$
\eqalign{
ds^2&=H^{{1\over3}}\big[ H^{-1} ds^2(\bE^{(1,2)})
+ ds^2(\bE^8)\big]
\cr
ds^2&=H^{{2\over3}}\big[ H^{-1} ds^2(\bE^{(1,5)})
+ ds^2(\bE^5)\big]\ ,}
\eqn\mtf
$$
respectively,
the metrics of the  D-p-branes [\hs, \dl] are
$$
ds^2= H^{-{1\over2}} ds^2(\bE^{(1,p)})
+H^{{1\over2}} ds^2(\bE^{9-p})
\eqn\dpppp
$$
the metric of the fundamental string [\dgh] is
$$
ds^2= H^{-1} ds^2(\bE^{(1,1)})+ ds^2(\bE^{8})
\eqn\nsss
$$
and the metric of the NS-5-brane [\chs] is
$$
ds^2=  ds^2(\bE^{(1,5)})+H ds^2(\bE^{4})\ ,
\eqn\nsf
$$
where $H$ is a harmonic function of the
 transverse directions in each case.
A common characteristic of all these 
solutions is that
the ratio $B^2A^{-2}$ of the 
conformal factor of the transverse
directions modulo that of the 
worldvolume directions is equal to
$H$, 
$$
B^2A^{-2}=H\ .
\eqn\abm
$$
In particular this implies that 
this ratio is invariant under
the various T- and S-dualities 
that relate the M-theory and
 type II string theories.
It also follows from the above 
observation and the harmonic 
function rule [\arkadyb] that for all the
intersecting brane 
configurations the ratio of the conformal
factor of the {\it overall transverse} directions 
modulo that of the {\it common intersection}
is universal and depends 
only on the number $n$ of the branes involved in the
intersection. In particular, we find that
$$
B^2A^{-2}=H_1 H_2 \cdots H_n\ .
\eqn\abp
$$
Incidentally, this ratio is  the
 same as that  of the spatial modulo the
timelike components of the metric of the associated black holes.

One expects that the moduli space of 
four- and five-dimensional
black holes, that arise from appropriately 
reducing the above brane solutions is flat. This is 
because the associated black holes preserve $1/2$
of the maximal supersymmetry and so 
the effective action has sixteen
supersymmetries. Such a high number of 
supersymmetries render the sigma model target
space
flat\foot{It also follows from the 
reduction of the effective
theories of branes to lower dimensions 
neglecting possible non-abelian
interactions and collecting the terms quadratic in 
the velocities.}. In such case, one can
choose for the black hole moduli potential
$$
\mu=\int d^Dx H\ .
\eqn\single
$$
Now if two or more  branes are 
involved in the configuration, it is natural to
take  the moduli potential to 
depend on the product of harmonic functions.
This is because all harmonic 
functions enter in a symmetric way in the
black hole metric\foot{There 
are completely symmetric brane intersections
that give rise to four- and 
five-dimensional black holes. For these
 the metric,  the scalars and 
 the Maxwell fields are all symmetric.}
  and that if one of them
is set to one, say
$H_k=1$, $1\leq k\leq n$, the 
expression for the moduli potential should
 remain symmetric in the rest of the
harmonic functions. Of course there 
are many other symmetric polynomials of the
harmonic functions which can be 
added in the expression for 
the potential. But all of them have
degree lower than that of
 the product. After setting one 
or more harmonic functions to one, we
get a moduli metric which would 
be scaled by a conformal factor. In particular,
we shall find a scaled version of 
the potential \single\ that is not expected. 

Another argument in support of 
\unipot\ can be established using duality.
As we have mentioned the
 expression for the moduli potential
is T- and S- dualities invariant.
 Now if we assume that black holes
that are related by T- and 
S-dualities should have the same moduli space,
then the moduli potential 
\unipot\ can be derived in the five-dimensional case.
This is because the moduli potential 
for the graviphoton black hole agrees with
\unipot\ and that this black hole 
is a reduction  from the 
M-theory configuration of
three M-2-branes 
intersecting on a 0-brane. The M-2-brane 
configuration  is then related to the
rest of intersecting branes of 
strings and M-theory via T- and S-dualities
which give the rest  of 
five-dimensional black holes.

It is worth mentioning that 
our expression for
the moduli potential \unipot\
 can apply to intersecting
brane configurations. In this 
case, the effective theory
may not be one-dimensional. Typically,
the dimension of the effective theory
is that of the common intersection. Moreover
the effective theory may contain apart from
scalars other fields like vectors and tensors.
However, we argue that the part of the effective
theory which describes
 the dynamics of the scalars that are associated
  with the overall
position of the configuration in spacetime when reduced
to one-dimension coincides with the effective
theory of the black hole that can be constructed
from the configuration.
To give an example, let us consider the moduli of the
 solution of eleven-dimensional supergravity with metric
$$
\eqalign{
ds^2&=H_1^{{1\over3}}H_2^{{1\over3}}H_3^{{1\over3}}
\big[-H_1^{-1}H_2^{-1}H_3^{-1}ds^2(\bE)
\cr &+H_1^{-1}ds^2(\bE^2)+H_2^{-1}ds^2(\bE^2)
+H_3^{-1}ds^2(\bE^2)+ds^2(\bE^3)\big]}
\eqn\inteintera
$$
which has the interpretation of 
three M-2-branes intersecting on
a 0-brane, where $H_1, H_2, H_3$
are harmonic functions on $\bE^3$. 
The effective theory of the transverse 
scalars this configuration
is determined by the moduli potential
$$
\mu=\int\, d^3x\, H_1 H_2 H_3
\eqn\moddddd
$$
which is the ratio of the component of the metric
along $\bE^3$ (overall transverse)
 to the component of the metric along 
 $\bE$ (common intersection).
The above moduli potential is of course
the moduli potential of the four-dimensional
black hole associated with the above configuration.
A similar analysis can be done for other such configurations.

Despite these there are some puzzles. 
One involves the M-theory
configuration of three M-5-branes 
pairwise intersecting on a 3-brane and all
together at a string [\gppkt]. Reducing
 this solution to five dimensions, we find
a string solution which is in the same 
universality class as the string solutions
of the N=2 D=5 supergravity. One might 
expect that the effective theory
of such strings using the supersymmetry 
projectors of the M-branes to be
a (4,0)-supersymmetric two-dimensional 
sigma model with strong HKT geometry.
This would imply that the moduli space
 has dimension $4N$ and that the torsion 
is a closed three form. Since there are three 
transverse scalars for each black hole,
the moduli space has in fact dimension
 $3N$ and the torsion is not a closed form.
The former point can be explained by 
arguing that there are additional moduli
for these black holes. Indeed in 
string theory apart from the transverse
scalars, intersecting brane configurations have 
additional scalar, vector and or even tensor
moduli. The additional scalars may be
 due to e.g.  D-brane type of counting;
for an example see [\john].
All these can be reduced to five
 dimensions giving a bigger moduli space from
the one we are investigating.
The latter point may also be resolved 
by observing that the intersection
is chiral. In such a case the torsion 
can be modified by adding a Chern-Simons
type of term to cancel the anomaly 
which renders the modified torsion to be
a non-closed form. However, we have not been able
to establish the details of the above arguments.
One encounters similar puzzles with even purely
electric solutions as it has already 
been mentioned in [\gary]. As another one
consider the seven-dimensional black 
hole which can be found by reducing
the intersection of two M-2-branes 
on a 0-brane configuration of M-theory.
One can easily see that in this 
case there is missing moduli. However
upon reducing the solution further 
to five-dimensions, the moduli space
becomes strong HKT as expected. This appears to be
a   rather
common phenomenon.  Upon reducing the 
solutions to an {\sl appropriate}  dimension,
the geometry of the associated black 
hole moduli space can be understood
in terms of that of the target space 
of a supersymmetric sigma model.

\chapter{Very special Four-Dimensional Black Holes}

A large class of black holes 
in four dimensions can be
constructed by reducing either 
the string solution of the
N=2 supergravity superposed with a pp-wave or the 
very special electrically charged
black holes of the N=2 D=5 supergravity in a 
KK-monopole background.
To describe these solutions, we first 
review some aspects of N=2 D=5 supergravity.
 The bosonic part  of the action of
five-dimensional
$N=2$ supergravity  with $k$ vector 
multiplets is associated to  a
 hypersurface $N$ of 
$\bR^k$ defined by the equation 
$$ 
V(X) \equiv {1 \over 6} C_{IJK}\, X^I X^J X^K=1 
\eqn\bone
$$ 
where $\{X^I; I=1, \dots, k\}$ are
 standard  coordinates 
on $\bR^k$ and $C_{IJK}$ are 
constants. In the case of a
model arising from a Calabi-Yau 
compactification of M-theory, the constants 
$C_{IJK}$ are the topological 
intersection numbers of the compact manifold.
Next we set
$$
\eqalign{
Q_{IJ}&\equiv-{1 \over 2}{\partial \over \partial X^I}
{\partial \over \partial
X^J} \log V \mid_{V=1}
={9 \over 2}X_I X_J - {1 \over 2}C_{IJK}X^K
\cr
h_{ab} &= Q_{IJ} {\partial X^I \over \partial \f^a}
{\partial X^J \over \partial
\f^b} \mid_{V=1}\ ,}
\eqn\btwo
$$
where $\{\phi^a; i=1,\dots, k-1\}$ are local
 coordinates of $N$, $h$ is interpreted as  a metric
on $N$ and
$$
X_I ={1 \over 6}C_{IJK}X^J X^K
\eqn\bthree
$$
are the dual coordinates to $X^I$. Note that 
the hypersurface equation $V=1$
can also be rewritten as  $X^I X_I =1$.
Then, the bosonic part of the associated 
supergravity  action [\town, \wit]
with vector potentials $A^I$ and scalars $\phi^a$ is  
$$
\eqalign{
S =& \int d^5 x \sqrt{-g} 
\big[ R +{1 \over 2}Q_{IJ}{F^I}_{\m
\n}{F^J}^{\m \n}
+h_{ab}\pd{\m} \f^a \pu{\m} \f^b\big] 
\cr &-{1 \over
24}e^{\m \n \p \s \t}C_{IJK}
{F^I}_{\m \n}{F^J}_{\p \s}{A^K}_\t\ ,}
\eqn\acti
$$
where $F^I = d A^I$, $I,J,K=1,\dots, k$ 
are the 2-form Maxwell field strengths,
$\mu, \nu, \rho, \sigma=0, \dots, 4$,
and $g$ is the metric
 of the five-dimensional spacetime;
we have used the same symbol 
$\phi^a$ to denote both
the coordinates of $N$ and the
 various scalar fields of the theory. 
 
The field equations of the above 
Lagrangian obtained from varying
the scalars $\f^a$, the spacetime
 metric $g$, and the vector
potentials $A^I$  are
$$
\eqalign{
\sqrt{-g} \partial_a Q_{IJ} 
\big[ {1 \over 2}{F^I}_{\m \n}{F^J}^{\m
\n}+\pd{\m}X^I \pu{\m}X^J \big] 
- 2 \pd{\m} \big( \sqrt{-g}Q_{IJ}
\pu{\m}X^I \big) \partial_a X^K =0\ ,}
\eqn\feqa
$$
$$
\eqalign{
\sqrt{-g} \big( G_{\m \n}
+Q_{IJ}{F^I}_{\m \p}{{F^J}_{\n}}^\p
+Q_{IJ} \pd{\m}X^I \pd{\n}X^J \big)
\cr
-{1 \over 2}\sqrt{-g}
g_{\m \n} \big[ {1 \over 2}Q_{IJ} {F^I}_{\p
\s}{F^J}^{\p \s} 
+Q_{IJ}\pd{\p}X^I \pu{\p}X^J \big]=0\ ,}
\eqn\feqb
$$
and
$$
\eqalign{
-2 \pd{\m} \big[ \sqrt{-g} 
Q_{IJ} {F^J}^{\m \n} \big]
 -{1 \over
8}e^{\n \p \s \m \t}
C_{IJK} {F^J}_{\p \s}{F^K}_{\m \t} =0\ ,}
\eqn\feqc
$$
respectively. The electrically 
charged black holes have been found in [\sabra].
The electrically charged
 black hole solutions in the background
of a KK-monopole are 
$$
\eqalign{
ds^2 &= -e^{-4U} dt^2 +e^{2U}
\big[ H_0^{-1} (d\tau+\omega)^2+H_0 d{\bf x}^2\big]
\cr
{A^I}_0 &= e^{-2U} X^I
\cr
e^{2U}X_I& ={1 \over 3}  H_I\ ,}
\eqn\sol
$$
where
$$
\eqalign{
H_I &= h_I + \sum_{A=1}^{N_I} {{\l_{IA}} \over
|{\bf{x}}-{\bf{y}}_{IA}|}
\cr
H_0&=h_0+\sum_{A=1}^{N_0}{{\l_{0A}} \over
|{\bf{x}}-{\bf{y}}_{0A}|}}
\eqn\ham
$$
are harmonic function on $\bR^3$. 
The positions of the black holes
are labeled  by the coordinates 
$\{({\bf{y}}_{0A},{\bf{y}}_{IA}); 
A=1, \dots, N_I; I=1, \dots, k\}$. 
The constants $\{(h_0, h_I); I=1, \dots, k\}$
are the values of the scalar 
fields at spatial infinity and 
$\{(\l_{0A}, \l_{IA}); A=1, \dots, N_I ; I=1, \dots, k\}$
 are the charges
of the black holes.   Viewing $e^U$
as an additional scalar, 
the last equation in \sol\
gives the  $k$ independent 
scalars $\{e^U, \phi^a\}$
 in terms of the $k$ harmonic
functions $\{H_I\}$. 
A reduction of this solution to 
four dimensions along the compact direction
$\tau$ leads to four-dimensional 
black holes with metric (in the Einstein frame)
$$
ds^2=-e^{-3U} H_0^{-{1\over2}} dt^2
+e^{3U} H_0^{{1\over2}} d{\bf x}^2\ .
\eqn\fone
$$
These black holes, apart from
 the electric charges associated
with those in five dimensions,  also have a magnetic
charge with respect to the KK-vector of the reduction.

The other class of four-dimensional 
black holes can be obtained by
using the string solution of N=2 
D=5 supergravity of [\sabratwo]. Superposing
this solutions with a pp-wave, we have
$$
\eqalign{
ds^2&= e^{-U} \big (du dv+ H^0 du^2\big)
+e^{2U} d{\bf x}^2
\cr
F^I_{mn}&=-\epsilon_{mnp} \partial_pH^I
\cr
e^U X^I&=H^I\ ,}
\eqn\hill
$$
where $u,v$ are light-cone coordinates and
$$
\eqalign{
H^I &= h^I + \sum_{A=1}^{N_I} {{\l^I_{A}} \over
|{\bf{x}}-{\bf{y}}_{IA}|}
\cr
H^0&=h^0+\sum_{A=1}^{N_0}{{\l^0_{A}} \over
|{\bf{x}}-{\bf{y}}_{0A}|}\ .}
\eqn\hama
$$
The positions of the black holes
are labeled by the coordinates 
$\{({\bf{y}}_{0A},{\bf{y}}_{IA}); 
A=1, \dots, N_I; I=1, \dots, k\}$ 
as in the previous case. 
The constants 
$\{(h^0, h^I); I=1, \dots, k\}$
are the values of the scalar 
fields at spatial infinity and 
$\{(\l^0{}_A, \l^I{}_A); 
A=1, \dots, N_I ; I=1, \dots, k\}$ are the charges
of the black holes.
Reducing this solution
 to four-dimensions along the 
direction of the wave
leads to a four-dimensional
black hole with metric in 
the Einstein frame (see also [\sabralust])
$$
ds^2=-H_0^{-{1\over2}} e^{-{3\over2}U}dt^2
+H_0^{{1\over2}} e^{{3\over2}U} d{\bf x}^2\ .
\eqn\ftwo
$$
These black holes carry magnetic
 charges which correspond to the
charges of the five-dimensional 
string. They also carry an electric
charge which is related to the 
momentum of the pp-wave.

Many other dyonic black hole solutions of
 four-dimensional supergravity theories are known.
Some of them have been found by investigating
the solutions of supergravity 
theories associated with the heterotic string [\mirone].
The relation of these black holes to brane configurations
of the heterotic string have also been explored 
[\mirtwo].

To express the black hole 
solutions explicitly in terms of the
harmonic functions, one has 
to solve the stabilization equations
(see [\kalloshc]). From here on, 
we shall assume that solutions
to these equations exist for 
the models we are considering.
We also remark that a special 
subclass of solutions are those
for which the positions of 
the harmonic functions are the same
 ${\bf y}_{0A}={\bf y}_{IA}={\bf y}_{JA}$
for $I\not= J$. These black
 holes are of interest 
because they exhibit regular 
horizons. It is straightforward
to see this by extending the 
arguments of [\cfgk, \sabratwo] which
have followed earlier work in 
[\gwgaa, \renatapeet, \gwgpktb].
In both cases the near 
horizon geometry is $AdS_2\times S^2$.


\chapter{The Moduli Potential of N=2 Black Holes}

The universal formula \unipot\ 
for the moduli potential is in 
agreement with the explicitly computed
moduli potential of the 
electrically charged black holes of N=2 D=5
supergravity with any number of 
vector multiplets [\gutpap]. Moreover, \unipot\
is also in agreement with the 
explicitly computed moduli potential 
of the Reissner-Nordstr\"om black hole in [\mss].

Applying \unipot, we find that the moduli 
potential for the black holes \fone\ is
$$
\mu_1=\int d^3x\, H_0 e^{6U}\ ,
\eqn\poto
$$
while for the black holes \ftwo\ is
$$
\mu_2=\int d^3x\, H^0 e^{3U}\ .
\eqn\potb
$$
If we again assume that the moduli 
metric is invariant under duality, then
in both cases the moduli potentials 
can be derived from
that which we  compute in section eight. 
This is because we have mentioned
that the above dyonic black holes are 
dual to the electrically charged
 ones of N=2 D=4 supergravity.

To give some examples, we have to 
consider models for which the
stabilization equations have 
an explicit solution. Such a model  is 
 the so called STU model [\stu]; for 
 other models see [\sabrac, \kalloshb].
  For this, the only non vanishing 
 component of $C_{IJK}$
is $C_{123}$.   For the back 
holes \fone, one can find that 
$$
e^{6U}= H_1 H_2 H_3\ .
\eqn\potc
$$
Therefore, the moduli potential of the 
associated four-dimensional black holes is
$$
\mu_1=\int d^3x\, H_0  H_1 H_2 H_3\ .
\eqn\potd
$$
Similarly for the \ftwo\ black holes, we find that
$$
e^{3U}= H^1 H^2 H^3\ .
\eqn\pote
$$
and consequently, the moduli potential is
$$
\mu_2=\int d^3x\, H^0  H^1 H^2 H^3\ .
\eqn\pof
$$

It is apparent that the moduli 
potential in both the above cases is
the same. So the geometry on the 
moduli space of \fone\
black holes is the same as that 
on the moduli space of \ftwo\ black holes.
Moreover, it coincides with the 
moduli potential of four-dimensional
black holes which preserve $1/8$
 of the maximal supersymmetry which
 are associated with 
 intersecting branes and have four 
 harmonic functions in sections two and three.
As for the intersecting 
branes black holes, we can set 
$H=H^0=H^1=H^2=H^3$
or $H=H_0=H_1=H_2=H_3$. 
Then the moduli potentials 
$\mu_1$ and $\mu_2$ reduce to 
that of Reissner-Nordstr\"om
black hole computed in [\mss].
A more general case is to take the 
positions of the harmonic 
functions of \fone\ black holes
to be the same but 
allow the charges $\l_{IA}$ and the 
asymptotic values of the scalars
to be different, i.e 
${\bf y}_A={\bf y}_{0A}
= {\bf y}_{IA}$ for $I=1,\dots, k$ 
but $h_0\not= h_I\not= h_J$
and $\l_{0A}\not=\l_{IA}\not=\l_{JA}$ 
for $I\not= J$ , and
similarly for the \ftwo\
 black holes. As we shall see,
this class of black holes 
exhibit the same superconformal 
properties for small black hole 
separation as those of the
Reissner-Nordstr\"om black hole.


\chapter{The effective theory of 
five-dimensional black holes}
\section{Supersymmetry}

It is straightforward given
 the moduli potential $\mu$,  \unipot, of 
the five-dimensional black holes to determine
the  metric and the torsion 
on the moduli space. 
This analysis is the same 
for all five-dimensional black holes,
i.e. those that have the 
interpretation as intersecting branes
in ten or eleven dimensions that 
preserve $1/8$ of the supersymmetry and the black
holes of N=2 D=5 supergravity that
preserve $1/2$ of the supersymmetry. So in what follows we
shall not distinguish between these two cases. 
The moduli space is a weak HKT 
manifold [\twist] with HKT potential $\mu$. 
So using [\poon, \michb], the metric and torsion are
$$
\eqalign{
ds^2&=\big[\partial_{mIA}\partial_{nJB}
+\sum_{s=1}^3 (I_s)^\ell{}_m (I_s)^q{}_n\,
\partial_{\ell IA}\partial_{qJB}\big]\mu\,
 dy^{mIA}\, dy^{nJB}
\cr
c&=6\partial_{pIA}\partial_{qJB}\partial_{sKC}\,\mu\, 
(I_1)^p{}_m (I_2)^q{}_n (I_3)^s{}_\ell 
dy^{mIA}\wedge dy^{nJB} \wedge dy^{\ell KC}} 
\eqn\mmetrica
$$
respectively, where $\{y^{mIA}; 
m=1, \dots, D-1; I=1,\dots, k;
A=1, \dots, N\}$ label
the positions of the $k N$ black-holes. 
The endomorphisms $\{I_r; r=1,2,3\}$
are associated with  a constant hypercomplex 
structure on $\bR^4$. These induce  a hypercomplex
structure on the moduli space by
setting
$$
({\bf I}_r)^{mIA}{}_{nJB}=
(I_r)^m{}_n \delta^I{}_J  \delta^A{}_B
\eqn\hype
$$
which is required for the HKT structure. 
One can easily show that the black hole
moduli space equipped with metric and 
torsion \mmetrica\ and hypercomplex structure
\hype\ admits an weak HKT 
structure\foot{For other applications of HKT
manifolds see [\tesc, \poon].}.

The effective theory has N=4B  
one-dimensional supersymmetry.
Both the supersymmetry multiplet 
and the effective action can be constructed
using the general results on supersymmetric sigma models of
 [\coles, \gary] adapted to this case.
In particular, we promote the 
coordinates on the moduli space to $N=4B$
superfields $y^{m IA}
=y^{m IA}(t, \theta^0, \dots, \theta^3)$ and impose 
  the
constraint
$$
D_r y^{m IA}= ({\bf I}_r)^{mIA}{}_{nJB} D_0y^{n IA}
= (I_r)^m{}_n D_0y^{n IA}
\eqn\pcon
$$
where  $\{ D_0, D_r ; r=1,2,3\}$ are 
the supersymmetry derivatives,
i.e.
$$
\eqalign{
D_0^2=D_r^2&=i\partial_t
\cr
D_0 D_r+D_r D_0&=0
\cr
D_r D_s+D_sD_r&=0, \qquad r\not=s\ .}
\eqn\susyd
$$
The associated  $N=4B$ supersymmetric action  is
$$
\eqalign{
S=-{1\over2}\int dt d\theta^0 \big[&ig_{mIA, nJB} 
D_0y^{m IA} \partial_ty^{n JB} 
\cr &+{1\over3!}c_{mIA, nJB, \ell KC}
D_0y^{m IA} D_0y^{nJB} D_0y^{\ell KC}\big]\ .}
\eqn\susyact
$$
This action describes the effective 
theory of five-dimensional supersymmetric
black holes which preserve four supercharges.

In the special case where the positions
 of the harmonic functions are the same,
${\bf y}^A={\bf y}^{IA}$,  the moduli 
space is again a weak HKT manifold
with HKT potential $\mu$. The metric 
and torsion are given by
$$
\eqalign{
ds^2&=\big[\partial_{mA}\partial_{nB}
+\sum_{s=1}^3 (I_s)^\ell{}_m (I_s)^q{}_n
\partial_{\ell A}\partial_{qB}\big]\mu\, dy^{mA}\, dy^{nB}
\cr
c&=6\partial_{pA}\partial_{qA}\partial_{sA}\,\mu\, 
(I_1)^p{}_m (I_2)^q{}_n (I_3)^s{}_\ell 
dy^{mA}\wedge dy^{nB} \wedge dy^{\ell C}\ ,}
\eqn\mmetricaa
$$
respectively. The hypercomplex
structure on the moduli space is
$$
({\bf I}_r)^{mA}{}_{nB}
=(I_r)^m{}_n   \delta^A{}_B\ .
\eqn\hypee
$$

The effective theory has again $N=4B$ 
supersymmetry one-dimensional supersymmetry.
Using again [\coles, \gary], the $N=4B$
superfields $y^{mA}=y^{mA}(t, \theta^0, \dots, \theta^3)$ 
 satisfy the
constraint 
$$
D_r y^{mA}= ({\bf I}_r)^{mA}{}_{nB} 
D_0y^{nA}= (I_r)^m{}_n D_0y^{nA}\ .
\eqn\mcones
$$
The  action of the effective theory is 
$$
S=-{1\over2}\int dt d\theta^0 \big[i 
g_{mA, nB} D_0y^{mA} \partial_ty^{nB}
 +{1\over 3!}c_{mA, nB, \ell C}
D_0y^{mA} D_0y^{nB} D_0y^{\ell C}\big]\ .
\eqn\eff
$$
This completes the description of the supersymmetric 
  effective theory actions of five-dimensional
black holes.

For black holes that preserve more 
than four supercharges, the moduli metric
is again determined by the moduli 
potential \unipot. However, the
effective theory may contain 
additional fermionic multiplets. For example,
for some black holes that 
preserve eight supercharges the moduli space
admits two commuting  strong 
HKT structures. The effective action then
contains additional fermions to 
construct the associated multiplets.
These multiplets have been 
described in [\coles, \gary].
Finally, we remark that the 
effective actions in both the above cases
 can also be written as a
full superspace integral as
$$
S=-{1\over2} \int\, dt\, d^4\theta\, \mu\ .
\eqn\anact
$$

\section{Superconformal Symmetry}

In [\mss], it was shown that for 
small black hole separations
the effective theory of the 
 graviphoton electrically charged black holes
of N=2 D=5 supergravity 
exhibits $D(2,1;0)$ superconformal symmetry.
This is related to the 
observation that the near horizon geometry
of these black holes is $AdS_2\times S^3$.
However, the near horizon 
geometry of five-dimensional black holes 
that preserve 1/8 of the maximal
supersymmetry and are associated 
with intersecting branes, and that 
of the black holes of the N=2 D=5 supergravity theory
is also $AdS_2\times S^3$.  Since 
this should be the case
for every black hole involved 
in the superposition, the relevant
solutions are those for which
 all the harmonic functions have the
same positions\foot{We also 
consider in our investigation only
STU black holes.}. 
So one expects that the effective 
theory \eff\ of these black holes will
also exhibit  $D(2,1;0)$ 
superconformal symmetry for small
black hole separations. 

The conditions for a $N=4B$ 
supersymmetric sigma model to exhibit
superconformal symmetry have 
been investigated in [\michb] and we shall
not repeat them here in detail\foot{Superconformal 
sigma models with scalar 
potential have been
considered in [\george].}. We 
shall simply verify the conditions
that the sigma model manifold 
admits a homothetic motion
generated by a vector field $D$ and
 that the associated one-form to $D$ is closed.
We shall then comment about the rest of the conditions.

The limit of small black hole 
separation is achieved  by requiring
that the asymptotic constants of
 the harmonic functions that determine the solutions
 vanish, i.e. $h_I\rightarrow 0$. Then
following [\micha], we write the moduli potential as
$$
\mu=\mu_1+\mu_2+\mu_3
\eqn\potslit
$$
where
$$
\mu_1=\int d^4x \sum_A {\lambda_{1A}\lambda_{2A}
\lambda_{3A}
\over |{\bf x}-{\bf y}_A|^6}
\eqn\post
$$
$$
\mu_2=\int  d^4x  \sum_{A\not=B}
 {\lambda_{1A}\lambda_{2A}\lambda_{3B}+
\lambda_{3A}\lambda_{2A}\lambda_{1B}
+\lambda_{1A}\lambda_{3A}\lambda_{2B}
\over |{\bf x}-{\bf y}_A|^4  |{\bf x}-{\bf y}_B|^2}
\eqn\poti
$$
$$
\mu_3=\int  d^4x  \sum_{A\not=B\not=C} 
{\lambda_{1A}\lambda_{2B}\lambda_{3C}
\over |{\bf x}-{\bf y}_A|^2  
|{\bf x}-{\bf y}_A|^2 |{\bf x}-{\bf y}_A|^2}\ .
\eqn\mpoti
$$
There is no contribution to the 
moduli metric  from $\mu_1$ because it
is independent from ${\bf y}_A$ 
as it can be easily seen
by a change of variables
 in the integral.
The contribution to the moduli 
metric due to $\mu_2$ has a 
logarithmic divergence
${{\rm ln}\delta \over |{\bf y}_A-{\bf y}_B|^2}$ 
for ${\bf x}\rightarrow {\bf y}_A$,
 where $\delta$ is a cut off, 
($|{\bf x}- {\bf y}_A|\geq \delta$. However,
these terms do not contribute 
to the moduli metric; they are eliminated
 passing from the potential to the moduli metric
  because of the differentiation. 
  The term of $\mu_2$ that contributes is
$$
\mu_2= 2\pi^2 \sum_{A\not=B} \big(\lambda_{1A}
\lambda_{2A}\lambda_{3B}+
\lambda_{3A}\lambda_{2A}\lambda_{1B}
+\lambda_{1A}\lambda_{3A}\lambda_{2B}\big) 
{{\rm ln}|{\bf y}_A-{\bf y}_B|
\over |{\bf y}_A-{\bf y}_B|^2}\ .
\eqn\mnpoti
$$
Finally, $\mu_3$ can be defined 
without regularization and 
it is homogeneous
of degree\foot{In the case for which all the positions
of the harmonic functions are 
different, $\mu_3$ is the only contributing
term. But as we shall see later, 
in this case the moduli metric is degenerate.} $-2$, i.e.
$$
y^{mA} \partial_{mA} \mu_3=-2  \mu_3\ .
\eqn\antipot
$$ 
The homothetic motion on the moduli space 
is generated by the vector field
$$
D^{mA} \partial_{mA}=-y^{mA} \partial_{mA}\ ,
\eqn\dilats
$$
which acting on the moduli metric gives
$$
{\cal L}_D g_{mA,nB}=2 g_{mA,nB}\ . 
\eqn\delot
$$
This can be verified by an 
explicit calculation using
the rotational invariance 
of the moduli potential
$$
\eqalign{
y^{mA} (I_r)^n{}_m\partial_{nA}\mu_2&=0
\cr
y^{mA} (I_r)^n{}_m\partial_{nA}\mu_3&=0\ .}
\eqn\rotinv
$$
Invariance of the effective 
action under special conformal
transformations requires that 
$$
D_{mA} dy^{mA}
\eqn\fdil
$$
is  a closed one-form, where we 
have used the moduli metric
to lower the indices of the 
components of $D$.
To show this, we first observe 
that the part of the metric
associated with $\mu_3$ is 
degenerate along $D$.
Using this, we find that
$$
\eqalign{
D_{mA} dy^{mA}&= -g_{mA, nB} y^{nB}dy^{mA}
\cr &
 =2\pi^2 d\big[ \sum_{A\not=B} 
 {\lambda_{1A}\lambda_{2A}\lambda_{3B}+
\lambda_{3A}\lambda_{2A}\lambda_{1B}
+\lambda_{1A}\lambda_{3A}\lambda_{2B} 
\over |{\bf y}_A-{\bf y}_B|^2}\big]\ ,}
\eqn\fidel
$$
and so $D_{mA} dy^{mA}$ is 
closed as required. It turns out that
the rest of the conditions for 
superconformal invariance also hold.
Therefore the moduli geometry 
for small black hole separation exhibits
a $D(2,1;0)$ superconformal invariance.


\chapter{The effective theory of
 four-dimensional black holes}
\section{Supersymmetry}

The effective theory of four-dimensional 
black holes which preserve
$1/8$ of the maximal supersymmetry 
is expected to have four supercharges.
The dimension of the moduli space 
is $3 n N$, where $n$ is the number
of harmonic functions of the solution and $N$ is the
number of positions of each harmonic function.
The description of the 
effective theory is the same for all 
four-dimensional black holes,
i.e. those that have the 
interpretation as intersecting branes
in ten or in eleven dimensions, 
the black holes that are reductions
of the solutions superposed
 with a pp-wave or a KK-monopole
of N=2 D=5 supergravity and
preserve 1/2 of the supersymmetry 
and the electrically
charged black holes of N=2 D=4 
supergravity. To keep 
the notation uniform, we label
the positions of the former black holes as 
$\{{\bf y}^{IA};
 I=1,\dots, n; A=1, \dots , N\}$ and 
the positions of the 
latter black holes as
 $\{{\bf y}^{IA}; 
 I=0,1,\dots, k; A=1, \dots , N\}$ with $n=k+1$.
The range of $I$ can be 
different in the two cases 
but this would not affect
our formulae below.

Next we derive the supersymmetry multiplet 
and the effective action of the above
black holes  by appropriately adapting
the general results of [\coles] on 
supersymmetric one-dimensional
sigma models and by comparing 
with the effective action
of Reissner-Nordstr\"om black
 holes in [\mss].
In particular, we promote the 
positions of the black holes to superfields as
${\bf y}^{IA}
={\bf y}^{IA}(t, \theta^0, \dots, \theta^3)$
 and add
a new supersymmetry fermionic 
multiplet $\psi^{IA}(t, \theta^0, \dots,
\theta^3)$. In addition, we impose the constraints
$$
\eqalign{
D_ry^{mIA}&=\epsilon_r{}^m{}_n D_0y^{nIA}
+\delta^m{}_r \psi^{IA}
\cr
D_r\psi^{IA}&=i\delta_{rn} \partial_ty^{nIA}\ .}
\eqn\concon
$$
We remark that all the four-dimensional 
black holes associated with
intersecting branes with four harmonic 
functions and those that are reductions of
N=2 D=5 supergravity are charged
 with respect one KK-vector. Therefore
the argument in [\mss] applies for the 
presence of the fermionic multiplets,
i.e. that they are due to zero modes 
along the KK-direction.
The manifestly supersymmetric 
effective action of the four-dimensional
black holes is
$$
S=-{1\over2} \int dt d^4\theta\, \mu(y)\ .
\eqn\esf
$$ 
Rewriting this action in terms of the N=1 superfields
$$
\eqalign{
q^{mIA}&=y^{mIA}|_{\theta^r=0}\ , \qquad r=1,2,3
\cr
\chi^{IA}&=\chi^{IA}|_{\theta^r=0}\ , \qquad r=1,2,3}
\eqn\fidelb
$$
by integrating over
 $\theta^1, \theta^2$ and $\theta^3$ and by using
the constraints \concon, we find
$$
\eqalign{
S=&\int dt d\theta \big[-{i\over2}
 g_{mIA,nJB} Dq^{mIA} \partial_tq^{nJB}-{1\over2}
h_{IA,JB} \chi^{IA} D \chi^{JB}
+i f_{mIA, JB} \partial_tq^{mIA} \chi^{JB}
\cr &
+{1\over3!} c_{mIA, nJB, \ell KC} 
Dq^{mIA}   Dq^{nJB} Dq^{\ell KC}
+{1\over2} n_{mIA, nJB, KC} Dq^{mIA}  
 Dq^{nJB} \chi^{KC}
\cr &+
{1\over2} m_{mIA, JB, KC}
 Dq^{mIA}   \chi^{JB} \chi^{KC}+
{1\over3!} l_{IA, JB, KC}
\chi^{IA} \chi^{JB} \chi^{KC}]\ , }
\eqn\effact
$$
where
$$
\eqalign{
g_{mIA,nJB}&=\big[\partial_{mIA}\partial_{nJB}
+\epsilon^{\ell p}{}_m  \epsilon_\ell{}^q{}_n 
\partial_{p IA}\partial_{qJB}\big]\mu\
\cr
h_{IA,JB}&=\delta^{mn} 
\partial_{mIA} \partial_{nJB}\mu
\cr
f_{mIA, JB}&=\epsilon^{n\ell}{}_m 
 \partial_{nIA} \partial_{\ell JB}\mu
\cr
c_{mIA, nJB, \ell KC}&={1\over2} 
\epsilon^{pqr} \epsilon_p{}^s{}_m
\epsilon_q{}^t{}_n 
\epsilon_r{}^u{}_\ell \partial_{sIA} \partial_{t JB}
\partial_{uKC}\mu
\cr
n_{mIA, nJB, KC}&=({1\over2} \epsilon^{pq\ell} 
\epsilon_p{}^s{}_m \epsilon_q{}^u{}_n+
\epsilon^s{}_m{}^\ell \delta^u{}_n) 
\partial_{sIA} \partial_{uJB} \partial_{\ell KC}\mu
\cr
m_{mIA, JB, KC}&={1\over2} 
\epsilon^{pqs} \epsilon_p{}^\ell{}_m
\partial_{\ell IA} 
\partial_{qJB} \partial_{sKC}\mu
\cr
l_{IA, JB, KC}&={1\over2} \epsilon^{mns} 
\partial_{mIA} \partial_{nJB} \partial_{sKC}\mu\ ,}
\eqn\condfil
$$
and we have set $\theta^0=\theta$.
In particular, the moduli metric is
$$
ds^2=\big[\partial_{mIA}\partial_{nJB}
+\epsilon^{\ell p}{}_m  \epsilon_\ell{}^q{}_n 
\partial_{p IA}\partial_{qJB}\big]\mu\, 
dy^{mIA}\, dy^{nJB}\ .
\eqn\mmetricb
$$

In the special case where the positions 
of the harmonic functions are the same,
${\bf y}^A={\bf y}^{IA}$,  to 
construct the effective action,
we again promote ${\bf y}^A$ to 
superfields and introduce $N$ additional
fermionic multiplets $\psi^A$. These
 multiplets satisfy the constraints
$$
\eqalign{
D_ry^{mA}&=\epsilon_r{}^m{}_n D_0y^{nA}
+\delta^m{}_r \psi^{A}
\cr
D_r\psi^{A}&=i\delta_{mn} \partial_ty^{nA}\ .}
\eqn\concona
$$
The effective action is again 
given by \esf. Expanding the action
in terms of the N=1 superfields
$$
\eqalign{
q^{mA}&=y^{mA}|_{\theta^r=0\ ,
 \qquad r=1,2,3}
\cr
\chi^{A}&=\chi^{A}|_{\theta^r=0}\, 
\qquad r=1,2,3\ ,}
\eqn\megn
$$
we find that
$$
\eqalign{
S=&\int dt d\theta \big[-{i\over2} 
g_{mA,nB} Dq^{mA} \partial_tq^{nB}-{1\over2}
h_{AB} \chi^{A} D \chi^{B}
+i f_{mA, B} \partial_tq^{mA} \chi^{B}
\cr &
+{1\over3!} c_{mA, nB, \ell C} 
Dq^{mA}   Dq^{nB} Dq^{\ell C}
+{1\over2} n_{mA, nB, C}
 Dq^{mA}   Dq^{nB} \chi^{C}
\cr &+
{1\over2} m_{mA, B, C} 
Dq^{mA}   \chi^{B} \chi^{C}+
{1\over3!} l_{ABC}\chi^{A} \chi^{B} \chi^{C}]\ , }
\eqn\effact
$$
where
$$
\eqalign{
g_{mA,nB}&=\big[\partial_{mA}\partial_{nB}
+\epsilon^{\ell p}{}_m 
 \epsilon_\ell{}^q{}_n 
\partial_{pA}\partial_{qB}\big]\mu\
\cr
h_{AB}&=\delta^{mn} 
\partial_{mA} \partial_{nB}\mu
\cr
f_{mA,B}&=\epsilon^{n\ell}{}_m  
\partial_{nA} \partial_{\ell B}\mu
\cr
c_{mA, nB, \ell C}&={1\over2}
 \epsilon^{pqr} \epsilon_p{}^s{}_m
\epsilon_q{}^t{}_n \epsilon_r{}^u{}_\ell
 \partial_{sA} \partial_{tB}
\partial_{uC}\mu
\cr
n_{mA, nB, C}&=({1\over2} \epsilon^{pq\ell} 
\epsilon_p{}^s{}_m \epsilon_q{}^u{}_n+
\epsilon^s{}_m{}^\ell \delta^u{}_n) 
\partial_{sA} \partial_{uB} \partial_{\ell C}\mu
\cr
m_{mA, B, C}&={1\over2} \epsilon^{pqs} 
\epsilon_p{}^\ell{}_m
\partial_{\ell A} \partial_{qB} \partial_{sC}\mu
\cr
l_{ABC}&={1\over2} \epsilon^{mns} 
\partial_{mA} \partial_{nB} \partial_{sC}\mu\ ,}
\eqn\zina
$$
and again we have set $\theta^0=\theta$.
In particular, the moduli metric is
$$
ds^2=\big[\partial_{mA}\partial_{nB}
+\epsilon^{\ell p}{}_m  \epsilon_\ell{}^q{}_n 
\partial_{pA}\partial_{qB}\big]\mu\, 
dy^{mA}\, dy^{nB}\ .
\eqn\mmetricbb
$$
This completes the description of the 
supersymmetric effective actions
for four-dimensional black holes.

For four-dimensional black holes 
that preserve more supersymmetry,
the moduli potential is again
 given by \unipot. However, in the description
of the effective theory one may 
have to add additional one-dimensional
fields to describe the 
supersymmetry multiplets. These are required
by supersymmetry as in the five-dimensional case.

\section{Superconformal Symmetry} 

The investigation of superconformal 
symmetry of the moduli space
of four-dimensional black holes
for small black hole separation 
is similar to that presented
for the Reissner-Nordstr\"om 
black hole in [\mss].
In particular, the effective 
theory admits a $D(2,1;0)$ superconformal
symmetry in the near horizon limit.
As in the four-dimensional case,
 the relevant class
of black holes are those that exhibit 
regular near horizon geometry.
These are the black holes that have 
four harmonic functions\foot{We
 also consider only the black holes
and string solutions of the N=2 
D=5 supergravity 
associated with the STU model.} 
with the same centres for which the
near horizon geometry at 
every centre is $AdS_2\times S^2$.

The conditions for a sigma model 
action such as   \effact\ to exhibit
superconformal symmetry have been 
given in [\mss] and we shall not repeat
them here. In what follows, we 
shall show that our moduli metric 
admits a homothetic motion 
generated by a vector field $D$ and that
the associated one-form of $D$ is closed. 
To begin, we write
$$
g_{mA, nB}= G^{k\ell}_{mn} 
\partial_{kA} \partial_{\ell B} \mu\ ,
\eqn\opmetric
$$
where
$$
G^{k\ell}_{mn}= \delta^k_m \delta^\ell_n
 +\epsilon^{rk}{}_m \epsilon_r{}^\ell{}_n\ .
\eqn\herc
$$ 
Since we expect  a close 
relationship between the
superconformal properties of 
the five-dimensional black holes
and those of the four-dimensional 
ones, we take  the vector field $D$
which generates the homothetic 
motion to be the following:
$$
D^{mA} \partial_{mA}={2\over h} 
y^{mA} \partial_{mA}\ ,
\eqn\zue
$$
where $h$ is a constant which will be determined.
Using our ansatz for $D$ and 
the expression \opmetric, we find that
$D$ is a homothety if
$$
 (y^{mA} \partial_{mA}-h)\mu={h\over2} K\ ,
 \eqn\conch
$$
where $K$ is in the kernel of the operator 
$$
G_{ABmn}=G^{k\ell}_{mn} 
\partial_{kA} \partial_{\ell B}
\eqn\diana
$$
which is used to find the  metric on the
moduli space from the moduli potential.

As in the case of five-dimensional 
black holes above, we write the
moduli potential as
$$
\mu=\mu_1+\mu_2+\mu'\ ,
\eqn\athena
$$
where
$$
\mu_1=\int d^3x \sum_A  
{\lambda_{1A}\lambda_{2A}\lambda_{3A}\lambda_{4A}
\over |{\bf x}-{\bf y}_A|^4}\ ,
\eqn\socrates
$$
$$
\mu_2=\int d^3x \sum_{A\not=B}\big [
{\lambda_{1A}\lambda_{2A}\lambda_{3A}\lambda_{4B}
\over |{\bf x}-{\bf y}_A|^3
|{\bf x}-{\bf y}_B|}
+{\rm cyclic\,\, in} (1,2,3,4)\big]\ ,
\eqn\plato
$$ 
and $\mu'$ contains the rest of the terms.
Both $\mu_1$ and $\mu_2$ 
contain divergent terms which however
do not contribute to the moduli 
metric. In particular,
 $\mu_1$ is independent form the 
positions of the black holes and 
so it does not contribute. 
Putting a cut off 
$|{\bf x}-\bf {y}_A|\geq\delta$, 
we can evaluate 
$\mu_2$ to find 
$$ 
\mu_2=4\pi\sum_{A\not=B}\big [ 
\lambda_{1A}\lambda_{2A}\lambda_{3A}\lambda_{4B}
{{\rm ln}|{\bf 
y_A}-{\bf y}_B|+ (1-{\rm ln}\delta)
\over |{\bf y}_A-{\bf 
y}_B|}+{\rm cyclic\,\, in} (1,2,3,4)\big]\ . 
\eqn\aristot
$$
Next using 
$$ 
G_{CDmn}{1\over |{\bf y_A}-{\bf y}_B|}=0\ , 
\eqn\archimedes
$$ 
we see that the 
divergent part does not contribute. 
So ignoring the divergent 
part, we see that 
$$ 
(y^{mA}\partial_{mA}+1)\mu_2=-{1\over2} K\ , 
\eqn\pluto
$$ 
where 
$$ K={\pi\over4} \sum_{A\not=B}\big [ 
\lambda_{1A}\lambda_{2A}\lambda_{3A}\lambda_{4B}
{1 \over |{\bf 
y}_A-{\bf y}_B|}+{\rm cyclic\,\, in} 
(1,2,3,4)\big]\ . 
\eqn\iasonas
$$ 
which is 
in the kernel of $G_{CDmn}$. Next, 
we can observe that $\mu'$ is 
homogeneous of degree $-1$ and so 
$$ 
(y^{mA}\partial_{mA}+1)\mu'=0\ . 
\eqn\olisses
$$ 
From all these, we find that 
\conch\ holds for $h=-1$ and 
$$ 
D=-2 y^{mA}\partial_{mA}
\eqn\taigetos
$$
is a homothetic vector field.
To generate special conformal 
transformations, the associated form
of $D$ should be closed as in the 
case of five-dimensional black holes.
In particular, one can show that
$$
D_{mA}=-2g_{mA, nB} y^{nB}=\partial_{mA}K\ .
\eqn\mont
$$
For this, we have used  the rotational 
invariance of $\mu$  under $SO(3)$,
i.e.
$$
y^{mA} \epsilon^{n\ell}{}_m \partial_{\ell B}\mu=0\ .
\eqn\lasttt
$$
The above properties of the moduli 
metric indicate
that at small black hole separation 
the effective theory  
admits an $SL(2,\bR)$
symmetry. It turns out that the rest 
of the conditions for superconformal
invariance of [\mss] can also be
 verified.  So the effective action
admits a $D(2,1;0)$ superconformal symmetry.


\chapter{Black
Holes and Special K\"ahler Geometry}
\section{The Black Hole Solutions}

The action of $N=2$ four-dimensional 
supergravity
 with $n+1$ vectors $F^I=dA^I$ and 
 $n$ complex scalars $z^a$
  has been found in  [\spgeo].
A class of such systems can be described in terms
of a holomorphic homogeneous 
of degree two potential
$F=F(X^I)$. It follows that
$$ \eqalign{ F&={1 \over 2} F_I X^I 
\cr F_I &= F_{IJ} X^J \cr X^I
F_{IJK}&=0 \cr 
X^I F_{IJKL}&= - F_{JKL}} \eqn\spb $$
where $F_I = {\partial \over \partial X^I}F$,
 $I=0,\dots, n$, and similarly for
higher order derivatives.
Next we set
$$ \eqalign{ e^{-K} \equiv i 
\big( {\bar{X}}^I F_I - X^I
{\bar{F}}_I \big) \cr N_{IJ} 
= i \big({\bar{F}}_{IJ} - F_{IJ}
\big) \cr \cN_{IJ} = {\bar{F}}_{IJ}
 +{i \over (XNX)}(NX)_I (NX)_J}
\eqn\spc 
$$
where $(NX)_I = N_{IJ} X^J$,$(N {\bar{X}})_I 
= N_{IJ}
{\bar{X}}^J$, $XNX= X^I X^J N_{IJ}$ and 
$ {\bar{X}} N {\bar{X}} =
{\bar{XNX}}$. The existence
of such a potential $F$ is not always 
guaranteed. The field equations
of $N=2$ four-dimensional
supergravity 
are invariant under symplectic
 reparametrizations. It
has been shown that one may use this to pass
from a solution which possesses a 
potential to one which does not.
However, throughout this section 
we shall limit ourselves to
configurations which possess a 
potential $F$. The coordinates $X^I$
are holomorphic functions of $z^a$. 
 The bosonic part of the   $N=2$ four dimensional
supergravity action is
$$ 
\eqalign{ S = \int\, d^4x\, \sqrt{|g|} \big(R
 +2 \big[ e^K N_{IJ}
+e^{2K} (N {\bar{X}} )_I (NX)_J \big]
 \pd{\m} X^I \pu{\m}
{\bar{X}}^J \cr +i ( \cN_{IJ}
 - {\bar{\cN}}_{IJ}) {F^I}_{\m \n}
{F^J}^{\m \n} \big)
 -{1 \over 2} ( \cN_{IJ} + {\bar{\cN}}_{IJ})
\e^{\m \n \p \s} {F^I}_{\m \n}{F^J}_{\p \s}\ ,} 
\eqn\spa 
$$
where we have chosen $\e^{0123}=+1$.
The field equations of \spa are as follows. 
The Einstein field equation is
$$
 \eqalign{ \sqrt{|g|} \big( G_{\m \n}
 + 2 \big[ e^K N_{IJ}
+e^{2K} (N {\bar{X}} )_I (NX)_J \big]
 \pd{(\m} X^I \pd{\n)}
{\bar{X}}^J \cr +2i ( \cN_{IJ}
 - {\bar{\cN}}_{IJ}) {F^I}_{\m \l}
{F^J}_{\n}^\l \big) 
\cr 
-{1 \over 2} \sqrt{|g|} g_{\m \n} \big( 2
\big[ e^K N_{IJ} 
+e^{2K} (N {\bar{X}} )_I (NX)_J \big]
 \pd{\l} X^I
\pu{\l} {\bar{X}}^J
 \cr
 +i ( \cN_{IJ} - {\bar{\cN}}_{IJ})
{F^I}_{\p \s} {F^J}^{\p \s} \big) 
=0\ ,} 
\eqn\spd 
$$
the vector potential field equations are
$$ 8 \pd{\m} \big( \sqrt{|g|}
 \big[ {\rm Im} \cN_{IJ} F^{J \ \m
\n} + {\rm Re} \cN_{IJ}
 {^\star}F^{J \ \m \n} \big] \big) =0\ ,
\eqn\spe $$
and the field equations of 
the scalars $z^a$ and $\bar z^a$ are
$$ \eqalign{ \lbrace -2 \pd{\m}
 \big( \sqrt{|g|} \big[ e^K N_{LJ}
+e^{2K} (N {\bar{X}})_L (NX)_J \big]
 \pu{\m} {\bar{X}}^J \big) \cr
+2 \sqrt{|g|} \pd{L} \big(  e^K N_{IJ} 
+e^{2K} (N {\bar{X}})_I
(NX)_J \big) \pd{\m}X^I \pu{\m} {\bar{X}}^J 
\cr
 +i \sqrt{|g|}
\pd{L} (\cN_{IJ} - 
{\bar{\cN}}_{IJ}) {F^I}_{\m \n}F^{J \ \m \n} 
\cr
-{1 \over 2} \pd{L} (\cN_{IJ} 
+ {\bar{\cN}}_{IJ}) \e^{\m \n \p
\s}{F^I}_{\m \n} 
{F^J}_{\p \s} \rbrace \pd{a} X^L =0\ ,}
 \eqn\spf 
$$
and
$$ \eqalign{ \lbrace -2 \pd{\m}
 \big( \sqrt{|g|} \big[ e^K N_{LJ}
+e^{2K} (N {\bar{X}})_J (NX)_L \big]
 \pu{\m} {{X}}^J \big) \cr +2
\sqrt{|g|} \pd{{\bar{L}}}
 \big(  e^K N_{IJ} +e^{2K} (N
{\bar{X}})_I (NX)_J \big) \pd{\m}X^I \pu{\m}
 {\bar{X}}^J \cr +i
\sqrt{|g|} \pd{{\bar{L}}} 
(\cN_{IJ} - {\bar{\cN}}_{IJ}) {F^I}_{\m
\n}F^{J \ \m \n} 
\cr 
-{1 \over 2} \pd{{\bar{L}}} (\cN_{IJ} +
{\bar{\cN}}_{IJ}) 
\e^{\m \n \p \s}{F^I}_{\m \n} {F^J}_{\p \s}
\rbrace
\pd{{\bar{a}}} {\bar{X}}^L =0\ ,} 
\eqn\spg 
$$
respectively. In the equations above, 
$\partial_a$ and $\pd{{\bar{a}}}$ denote
 the partial
derivatives with respect to 
$z^a$ and $\bar z^a$, respectively.

 The static black hole
solution which we shall consider has
 $X^I$ real. In which case, $F$ 
 and all its derivatives can be chosen
 to be
purely imaginary. The solution is
$$
\eqalign{ 
ds^2 &= -e^K dt^2 +e^{-K} d {\bf{x}}^2
\cr
 {A^I}_0 &= e^K X^I 
 \cr {A^I}_m &= 0
 \cr
 2i F_I &= H_I}
  \eqn\spi
$$
where
$$ H_I = h_I + \sum_{A} {\l_{IA} 
\over |{\bf{x}} -
{\bf{y}}_{IA}|}\ . \eqn\spl $$
are harmonic functions. Using
 the definition of $K$, we 
find that for this solution $e^{-K}=4iF$.
Now the last equation in \spi\
 can be thought of as expressing $n+1$ real
scalars in terms of $n+1$ real harmonic 
functions. From these, $n$ scalars
are associated with the scalar
 fields of the supergravity theory
and one with the components of the 
metric as in the five-dimensional
case explained in section four. 
The centres of the harmonic functions
$\{{\bf{y}}_{IA}; I=0,\dots, n; 
A=1,\dots, n_I \}$ determine
 the positions 
of the black holes and 
$\{\l_{IA}; I=0,\dots, n; 
A=1,\dots, n_I \}$ are their
electric charges.
The above solution has 
delta function sources in the
coordinate system that 
we are using to describe it.
The appropriate source
 terms which should be added
to the supergravity action are 
$$ 
S_{source} = \sum_{I,A} \int\, d \t_{IA}\,
 (8 \pi e^{{1 \over 2}K}
(X^I + {\bar{X}}^I) \l_{IA} - 
16 \pi {A^I}_\m {d {y_{IA}}^\m \over
d \t_{IA}} \l_{IA})\ , 
\eqn\spm 
$$
where $\t_{IA}$ is the proper time 
associated with the 
centres ${\bf{y}}_{IA}$  defined
with respect to the metric $g$.

\section{Perturbations}

In order to determine the low 
energy behaviour of these solutions 
we allow the centres ${\bf{y}}_{IA}$ 
to depend on $t$. We also 
make the following additional first 
order in the velocities perturbations to the 
fields 
$$ \eqalign{ ds^2 &\ \rightarrow ds^2 
+ 2 e^K p_n dt dx^n 
\cr 
{A^I}_0 dt &\ \rightarrow {A^I}_0 dt 
+ ({D^I}_n - e^K X^I 
p_n)dx^n \cr X^I &\ \rightarrow X^I + i Y^I} 
\eqn\spn $$ 
where $Y^I$ is  real, and $p_n$, 
${D^I}_n$ ($n=1,2,3$) and $Y^I$ are to be 
determined by solving the 
equations of motion up to first order in 
the velocities \foot{One can 
perturb all of the fields in the theory
around a solution, but in this 
case the perturbation we have considered
will suffice.}. It turns out 
that using this perturbation
 ansatz only the $n$-component of the 
gauge equation and the $0n$ 
component of the Einstein equation 
together with the field equations of the scalar 
$z^a$ are modified by terms first order 
in the velocities. As the 
scalar perturbation is imaginary, the 
conjugate scalar equation
 does not contain any additional 
information. 

In particular the perturbed Einstein 
equation including the sources from \spm\ gives
$$
\eqalign{ -{1 \over 2} e^K \pu{\ell}
 (\pd{n}p_\ell - \pd{\ell}p_n)
+2X^I \pd{0}\pd{n}H_I \cr
 +4ie^K \pu{\ell}F_I
(\pd{n}{D^I}_{\ell}-\pd{\ell}{D^I}_n) 
= 8 \pi X^I \sum \l_{IA} \d
({\bf{x}}-{\bf{y}}_{IA}) v_{IAn}} 
\eqn\spo
$$
and the perturbed gauge equation 
including the sources gives
$$
\eqalign{ 4 \pd{0} \pu{n} H_I 
+8i \pd{m} \big( e^K
(F_{IJ}-F^{-1}F_I F_J) (\pu{m}{D^I}^n 
- \pu{n}{D^I}^m) \big) \cr
 +8i \pd{m} \big(e^{2K}F_I
(\pu{m}p^n - \pu{n}p^m) \big) 
-2 \e^{mm \ell} \pd{m}Y^L
\pd{\ell}(F^{-1}F_{IL}) \cr
 -2 \e^{m n \ell}
\pd{m} (F^{-2}Y^L F_I) \pd{\ell}F_L 
= 16 \pi \sum \l_{IA} \d
({\bf{x}}-{\bf{y}}_{IA}) v_{IA}^n.} 
\eqn\spp
$$

To proceed we contract \spp\ with  
${1 \over 2}X^I$ and subtract 
it from \spo. This leads to the
 simpler expression
$$
\eqalign{ \pu{m} \big[ {3i \over 8}F^{-2} 
(\pd{m}p_n - \pd{n}p_m)
+F^{-2}F_I (\pd{m}{D^I}_n -\pd{n}{D^I}_m) 
\cr 
+2F^{-{1 \over
2}}F_J {\e_{nm}}^{\ell} \pd{\ell} 
(F^{-{3 \over 2}}Y^J) \big]=0.}
\eqn\spq
$$
The perturbed scalar equation 
including sources gives
$$
\eqalign{ \pd{a} X^L \lbrace 
\pd{m}((-iF^{-1}F_{LJ}+{1 \over
2}F^{-2}F_L F_J ) \pu{m}Y^J) 
\cr
 +{i \over 2} F^{-2}
  \pu{m}Y^J (F_J \pd{m}F_L - F_L
\pd{m}F_J) \cr -{3 i \over 2} 
F^{-2}\pu{m}F_J \pd{m}F_L - {1 \over
2}Y^J \pd{m} (F^{-2} F_L \pu{m}F_J) 
\cr 
-{i \over 2}\e^{mnr}
(\pd{m}(F^{-1} F_{JL})
 +F^{-2}F_J \pd{m}F_L) (\pd{n}{D^J}_r -
\pd{r}{D^J}_n)
 \cr 
+{1 \over 4}F^{-{3 \over 2}} \e^{mnr} \pd{m}
(F^{-{1 \over 2}}F_L) 
(\pd{n}p_r - \pd{r}p_n) \rbrace  =0.}
\eqn\spr
$$

We shall not continue to present the 
solutions to these second order
(with respect to spatial derivatives) 
equations in this section,
instead we shall first evaluate the 
term in the action which is quadratic in 
the velocities. From there it will
 become clear that the perturbations solve
a set of first order equations.

\section{The Moduli Metric}

To compute the moduli metric, we 
must substitute  the
solution to the perturbed field 
equations found in the previous section
into the total action (including source terms) 
and compute the
part which is second order in velocities. It is 
expected, as a consequence of the BPS
condition, that the zeroth
and first order contributions to
 the action vanish. Substituting the 
perturbation ansatz into the action 
(including the sources), and
 collecting the terms quadratic 
in the velocities, we find
$$
\eqalign{ S^{(2)} =  \int d^4x \big[ -16
 \pd{0}F_I \pd{0} (FX^I) -8 \pi \sum e^{-K}
X^I \l_{IA} |{\bf{v}}_{IA}|^2 
\d ({\bf{x}}-{\bf{y}}_{IA}) \cr -{1
\over 4}e^{2K} (\pd{m}p_n 
- \pd{n}p_m)(\pu{m}p^n - \pu{n}p^m) \cr
-2 (\pu{m}D^{In}
-\pu{n}D^{Im})(\pd{m}k_{In}-\pd{n}k_{Im}) \cr
+2ie^{K} (-F_{IJ}
+F^{-1}F_I F_J) (\pd{m}{D^I}_n - \pd{n}{D^I}_m
\cr -e^K X^I (\pd{m}p_n 
- \pd{n}p_m)) (\pu{m}D^{Jn} - \pu{n}D^{Jm}
-e^K X^J (\pu{m}p^n - \pu{n}p^m)) 
\cr 
+({1 \over 2}F^{-2}F_I F_J -
F^{-1} F_{IJ}) \pd{m}Y^I \pu{m}Y^J 
\cr 
-F^{-2} Y^I \pu{m}Y^J F_J
\pd{m}F_I 
+{3 \over 2}F^{-2} Y^K Y^L \pd{m}F_K \pu{m}F_L 
\cr 
+Y^L
\e^{mnr} (\pd{m}(F^{-1}F_{JL})
+F^{-2}F_J \pd{m}F_L) (\pd{n}{D^J}_r
- \pd{r}{D^J}_n) 
\cr
 +{i \over 2}F^{-{3 \over 2}} Y^L \e^{mnr}
\pd{m} (F^{-{1 \over 2}}F_L) (\pd{n}p_r 
- \pd{r}p_n) \big]\ ,}
 \eqn\sps
$$
where 
$$ {k_I}^n = 
\sum {\l_{IA} {v_{IA}}^n \over
|{\bf{x}}-{\bf{y}}_{IA}|}\ . 
\eqn\spt $$
It should be noted that on varying 
the fields $p_n$, ${D^I}_n$ and
$Y^I$ one obtains the first order 
in the velocities field equations of the
previous section. To simplify the action we set
$$
\eqalign{ Q_{mn} & = \pd{m}p_n
 - \pd{n}p_m -{4 \over 3}H_I
(\pd{m}{D^I}_n - \pd{n}{D^I}_m)
 \cr & 
 +{4i \over 3}F^{3 \over 2}F_L
{\e_{mn}}^r \pd{r}(F^{-{3 \over 2}}Y^L)
 -4i F^{1 \over 2}Y^L
{\e_{mn}}^r \pd{r} (F^{-{1 \over 2}}F_L)  
\cr {Q^I}_{mn} & = \pd{m}{D^I}_n -
\pd{n}{D^I}_m 
+2 B^{IJ} (\pd{m}k_{Jn}-\pd{n}k_{Jm})
+{\e_{mn}}^{\ell} \pd{\ell}Y^I} \eqn\spu
$$
where
$$
 B_{IJ} = F^{-1} (F_{IJ}-F^{-1}F_I F_J) \ ;
\eqn\spxad
$$
$B^{IJ}$ is the inverse of the 
matrix $B_{IJ}$. We shall
assume that this inverse exists, 
this is certainly true in many
interesting cases, like those of 
intersecting D-branes. Remarkably then, the
second order action simplifies to
$$
\eqalign{ S^{(2)}=\int d^4x \big[
 -16 \pd{0}F_I \pd{0} (FX^I)
  -8 \pi \sum
e^{-K} X^I \l_{IA} |{\bf{v}}_{IA}|^2 
\d ({\bf{x}}-{\bf{y}}_{IA})
\cr +2 B^{IJ} (\pd{m}k_{In}
- \pd{n}k_{Im})(\pu{m}{k_J}^n -
\pu{n}{k_J}^m)
 \cr
 +{3 \over 4}e^{2K}Q_{mn}Q^{mn} -{1 \over
2}B_{IJ} {Q^I}_{mn}Q^{J \ mn}\big]}. 
\eqn\spv
$$
To proceed, we shall take $Q=0$ and 
$Q^I=0$. It turns out that these conditions
are sufficient for solving the
 perturbed field equations.
Furthermore, the portion of the above 
action which is independent of $Q$ and $Q^I$ 
is precisely that which leads to
 the effective action for the
black holes which possesses the expected
 $N=4$ worldline supersymmetry. 
 Solving $Q=Q^I=0$, we find
$$
\eqalign{ Y^I & ={1 \over 4 \pi} 
\int d^3 {\bf{z}} {1 \over
|{\bf{x}}-{\bf{z}}|} \e^{\ell m n} \pd{\ell}
(B^{IJ}(\pd{m}k_{Jn}-\pd{n}k_{Jm}))({\bf{z}})
 \cr {D^I}_n & = {1 \over 2 \pi}
\int d^3 {\bf{z}} 
{1 \over |{\bf{x}}-{\bf{z}}|} \pu{\ell}
(B^{IJ}(\pd{\ell}k_{Jn}
-\pd{n}k_{J \ell}))({\bf{z}}) \cr p_n & ={1 \over 4
\pi} \int d^3 {\bf{z}} 
{1 \over |{\bf{x}}-{\bf{z}}|} \pu{\ell}
(16iF X^I (\pd{\ell}k_{In}- \pd{n} k_{I \ell}) 
+4i {\e_{\ell n}}^r
(F_L \pd{r}Y^L - Y^L \pd{r} F_L))({\bf{z}}).} 
\eqn\spw
$$
Substituting the solutions for the 
perturbations back into $S^{(2)}$, we find
that
$$
\eqalign{ S^{(2)} =\int d^4x \big[
 -16 \pd{0}F_I \pd{0} (FX^I) 
 -8 \pi \sum
e^{-K} X^I \l_{IA} |{\bf{v}}_{IA}|^2 
\d ({\bf{x}}-{\bf{y}}_{IA})
\cr +2 B^{IJ} (\pd{m}k_{In}
- \pd{n}k_{Im})(\pu{m}{k_J}^n -
\pu{n}{k_J}^m) \big].}
 \eqn\spx
$$
The moduli metric can be read from $S^{(2)}$.
To analyze the geometry of the 
moduli space we use the identities
$$
\eqalign{ B_{IJ} X^J = {1 \over 3}
F^{-1}F_I \cr B_{IJ} \pd{\m}
(FX^J) = \pd{\m} F_I.} \eqn\spy
$$
It then follows that
$$ \pd{mIA} F = {i \over 2 }X^I \pd{m} 
\big[ {\l_{IA} \over
|{\bf{x}}-{\bf{y}}_{IA}|} \big]
 \eqn\spz
  $$
and
$$
\eqalign{ \pd{mIA}\pd{nJB} F^2 
=-{1 \over 4}e^{-K}X^I \d_{IJ}
\d_{AB} \pd{m} \pd{n} \big[ {\l_{IA} \over
|{\bf{x}}-{\bf{y}}_{IA}|} \big] \cr 
-{1 \over 2} B^{IJ} \pd{m}
\big[ {\l_{IA} \over |{\bf{x}}-{\bf{y}}_{IA}|}
 \big] \pd{n} \big[
{\l_{JB} 
\over |{\bf{x}}-{\bf{y}}_{JB}|} \big] \ ,} 
\eqn\spaa
$$
where there is no sum over I or J and 
$\pd{mIA}$ denotes partial differentiation
with respect to $y_{mIA}$. Using all 
the above we find
$$
S^{(2)} = \int\, dt\, {1 \over 2} 
g_{mIA \ nJB} {v_{IA}}^m {v_{JB}}^n.
\eqn\spdd\
$$
where
$$ g_{mIA \ ,  nJB} =\pd{mIA} \pd{nJB} \mu
 - \pd{nIA} \pd{mJB} \mu +
\d_{mn} \d^{r \ell} \pd{r IA} \pd{\ell JB}\mu 
\eqn\spbb $$
and
$$ \mu = -16 \int \, d^3x \, F^2 \ . 
\eqn\spcc $$
So the moduli metric is 
$$
ds^2 =   g_{mIA \ ,  nJB} dy^{mIA} dy^{nJB} \ .
\eqn\spcx
$$

As an example we may consider the STU model 
with $n=4$. For this case the potential
function $F$ is given by
$$
F(X^I)=i(X^0 X^1 X^2 X^3)^{1 \over 2} \ .
\eqn\spcxx
$$
This leads to the moduli potential 
$$
\m = 16 \int \, d^3x \, H_0 H_1 H_2 H_3 \ .
\eqn\spcxxa
$$
From this we observe that the moduli 
space metric possesses the expected $N=4$
supersymmetry together with the 
appropriate superconformal symmetry
for those black holes associated 
with harmonic functions with
the same centres. The moduli space metric
is generated by a potential function in 
agreement with the conjecture \unipot.

\chapter{Concluding Remarks}

We have proposed that the moduli metric of a 
large class of four- and five-dimensional
black holes can be determined by a moduli 
potential. In turn this can
be determined by the components of the 
metric of the black hole
solution as in \unipot.
Then we have provided evidence that such 
choice gives consistent results.
In particular, it describes all  the 
black hole moduli metrics  that have been computed
explicitly. In some cases, one can 
argue under certain assumptions that
the expression for the moduli potential
\unipot\ can be shown using duality. 
Moreover, the associated effective theories
of black holes which are constructed using 
\unipot\ exhibit the expected
superconformal behaviour at small black hole separations.

One can  extend our construction
 to find a  U-duality
invariant expression for the 
moduli potential. However, for this to be
consistent  one should use 
U-duality invariant black hole solutions [\mirjam];
for recent work see for example [\fre].
In four dimensions, such black
 holes carry the charges of all branes\foot{
Note however that in string theory [\pol] as well 
as in supergravity [\costa], D6-branes repel 
D0-branes; although this may change
if other brane charges are present.}.
It would be interesting to see 
whether our moduli potential formula
still applies in this case.

It is clear from our results
that the effective theories of black holes
that have regular horizons exhibit 
superconformal symmetry 
at small black holes separations.
This can prove useful in  understanding 
 multi-black hole quantum mechanics
using the suggestion of [\fubini] 
adapted for black holes [\peet, \ggpkt] 
and black hole moduli spaces 
[\micha, \michb, \mss].

\vskip 0.5cm
{\bf Acknowledgments:} We thank   A. Strominger for
 helpful suggestions
and comments. G.P. also thanks G.W. Gibbons for many
discussions on four-dimensional black hole moduli spaces. 
J.G. thanks EPSRC for a studentship. G.P. is
supported by a University Research
 Fellowship from the Royal Society.
 This work is partly
 supported by the PPARC grant  PPA/G/S/1998/00613.

\refout

\end